\documentclass[conference]{IEEEtran}
\IEEEoverridecommandlockouts

\usepackage{cite}
\usepackage{amsmath,amssymb}
\usepackage{graphicx}
\usepackage{booktabs}
\usepackage{multirow}
\usepackage{array}
\usepackage{url}
\usepackage{balance}
\usepackage{xcolor}
\usepackage[normalem]{ulem}
\definecolor{customcolor}{RGB}{0, 0, 238}
\usepackage[colorlinks=true, linkcolor=black, urlcolor=customcolor, citecolor=black]{hyperref}
\usepackage{todonotes}
\usepackage{orcidlink}
\graphicspath{{figures/}}

\title{Do Benign-Trained Hidden-State Probes Detect Tool Errors Under Adversarial Conditions?}

\author{
\IEEEauthorblockN{Bibek Bhandari$^{*}$\,\orcidlink{0009-0004-0491-4114}}
\IEEEauthorblockA{
Department of Computer Science and Engineering\\
CMR Institute of Technology, Bengaluru, India\\
bibekbhandari9848@gmail.com
}
\and
\IEEEauthorblockN{Kshitij Lingthep$^{*}$\,\orcidlink{0009-0005-5648-6737}}
\IEEEauthorblockA{
Independent Researcher\\
Memphis, Tennessee, USA\\
kshitij.lingthep.cs@outlook.com
}
\thanks{$^*$Both authors contributed equally to this work.}
}
\IEEEaftertitletext{\vspace{-2.0em}}
\newcommand{\citep}[1]{\cite{#1}}
\newcommand{\citet}[1]{\cite{#1}}
\newcommand{\isIEEE}{}

\usepackage{csquotes}
\MakeOuterQuote{"}
\begin{document}
\bstctlcite{IEEEexample:BSTcontrol}
\raggedbottom
\maketitle
\begin{abstract}
Although hidden-state probes efficiently monitor model operation, it remains unclear whether probes trained on benign tool use remain useful for detecting incorrect actions under adversarial conditions. In this research, we assess this problem from the perspective of operational tool-action correctness. we use four AgentDojo suites, four conditions, and three models: Qwen2.5-7B-Instruct, Qwen2.5-14B-Instruct, and GLM-4.6V-Flash (9B). We perform a five-fold, task-grouped out-of-fold (OOF) evaluation for each model. We train step-specific logistic-regression probes only on benign and non-defended trajectories and evaluate them on held-out trajectories for benign, prompt-injection, system-override, and tool-poisoning conditions. We compare our hidden-state probe to mean tool-call-token entropy, negative mean tool-call-token log-probability, and self-consistency uncertainty. The paired tests highlight six significant differences after Holm correction. All of the tests involve Qwen2.5-14B-Instruct under the system-override condition, with all three scalar uncertainty metrics obtaining higher area under the receiver operating characteristic curve (AUROC) and average precision (AP) than the hidden-state probe. These results suggest that benign-trained hidden-state probes should not be assumed to reliably detect operational tool-action errors under adversarial conditions.
\end{abstract}


\section{Introduction}
Nowadays, LLM agents increasingly engage with external tools to gather data and execute commands. This capability raises a reliability issue distinct from factual accuracy in regular text generation: an agent can issue an invalid command to a tool. Under adversarial conditions, untrusted tool outputs or conflicting instructions can further distort the agent's operational decisions \citep{greshake2023indirect,zhan2024injecagent,zhang2025asb}. Technologies like ReAct and Toolformer are demonstrations of the ways in which language models integrate reasoning and tool usage \citep{yao2023react,schick2023toolformer}.

Hidden-state probes use internal model representations to predict properties of model behavior. Recent research has investigated probes for improper tool choice and argumentation issues \citep{healy2026internal,yeats2026calls}. A parallel body of research focuses on whether the agent encodes indirect prompt injection at the level of representation, but the important caveat is that such recognition does not equate to safety \citep{dong2026latent}. Given these findings, we ask: can a hidden-state probe for predicting operational action correctness still be effective in an adversarial setting?

This matters because detecting an attack or injection surface is not the same as detecting an erroneous action. However, some recent studies on hidden-state probing have cautioned that discrimination may emerge from nuisance cues if the discrimination test does not account for the confounding variables being controlled. As such, we compare the benign-trained probe with uncertainty-based metrics using held-out task groups. This comparison builds on previous research on predictive uncertainty, self-consistency, and confidence calibration \citep{xiao2021uncertainty,farquhar2024semantic,manakul2023selfcheckgpt,wang2023selfconsistency}.

This paper audits and evaluates probes trained only on benign conditions for their ability to detect errors across three adversarial conditions. We compare the results with mean tool-call-token entropy, negative mean tool-call-token log-probability, and self-consistency uncertainty using task-grouped OOF predictions. We report the handling of exhaustion sentinels, missing tool-call-token boundaries, and pooled-step fallback in the evaluation population. We evaluate two Qwen model sizes alongside the GLM-4.6V-Flash model.

\section{Related Work}
\subsection{Tool-Using Agents and Benchmarks}
Early research set the stage for AI models that can actively interact with their environments by reasoning through problems step-by-step and learning to use external APIs to solve tasks \citep{yao2023react,schick2023toolformer}. Since then, agent evaluation has evolved from static questions into interactive scenarios involving tools and human workflows \citep{liu2024agentbench,ruan2024toolemu,yao2025taubench}. We use AgentDojo because of its realistic tasks and adversarial tool-output setting \citep{debenedetti2024agentdojo}.

\subsection{Prompt injection and agent security}
Indirect prompt injection occurs when an AI is hijacked by a malicious instruction hidden in the data it reads, and it is a practical risk for LLM-integrated applications \citep{greshake2023indirect}. To tackle this, researchers have built dedicated benchmarks and broader attack-defense evaluations for tool-integrated agents such as InjecAgent and Agent Security Bench \citep{zhan2024injecagent,zhang2025asb}.

\subsection{Tool-use error detection}
Earlier work showed that a model’s hidden activations contain enough signal to classify the truthfulness of its statements \citep{azaria2023internal}, and subsequent work explored internal-state features for hallucination detection \citep{chen2024inside}. Recently, Healy et al. used internal representations to identify tool-selection and argument-level errors \citep{healy2026internal}. Yeats et al. evaluate linear probes across a substantially broader set of tool-calling models and error types \citep{yeats2026calls}. These model-specific applications should be interpreted in light of general methodological concerns about what a probe learns from representations \citep{belinkov2022probing,hewittliang2019control}. The open question is whether a probe trained on benign action-correctness labels can distinguish incorrect tool actions under adversarial conditions.

\subsection{Prompt-injection exposure and probe validity}
While hidden states of LLMs can encode signals of indirect prompt-injection exposure in their hidden states,  it does not necessarily prevent them from following the injected instructions \citep{dong2026latent}. Li et al. cautioned that background cues can inflate the apparent accuracy of hidden-state probes \citep{li2026auc}. This highlights a broader issue with measuring an AI’s internal uncertainty. Whether we look at token-level uncertainty \citep{xiao2021uncertainty,farquhar2024semantic} or multi-sample consistency \citep{manakul2023selfcheckgpt,wang2023selfconsistency}, these uncertainty indicators are only useful when validated against ground truth. These studies isolate exposure detection from ground truth execution and place a strong emphasis on reporting what the label actually represents. Ultimately, our evaluation targets the correctness of the emitted tool call actions compared to the ground truth references, moving beyond a binary injection detection label. 

\subsection{Multi-step agent evaluation}
AgentHallu studies the attribution of hallucination-responsible steps in multi-step agent trajectories \citep{liu2026agenthallu}. This motivates a distinction between identifying the first point of divergence in a trajectory and evaluating the correctness of each emitted action. Our evaluation is narrower than general agent hallucination attribution: we examine step-level operational tool-action correctness in the available AgentDojo trajectories.
\section{Task Formulation and Detector Signals}
\subsection{Operational Action Correctness}

For a trajectory, let each benchmark-aligned step be indexed by task group $g$ and subsequent step $t$. Let $R_{g,t}$ denote the reference tool-call list and $A_{g,t}$ denote the model-generated tool-call list for that step. Either list may be empty. We assign a binary error label $y_{g,t} \in \{0,1\}$, where 0 indicates a correct step, and 1 indicates an incorrect step. The label is defined as:

\begin{equation}
    y_{g,t} =
    \begin{cases}
        0, & \operatorname{Match}_{\mathrm{bench}}(A_{g,t}, R_{g,t}), \\
        1, & \text{otherwise}.
    \end{cases}
\end{equation}

Here, $\operatorname{Match}_{\mathrm{bench}}$ indicates that the generated tool calls meet the benchmark's correctness requirements relative to the ground truth. A successful match includes the number and ordering of calls, tool selection, arguments, schema validity, and protocol constraints. 

However, incorrect steps include tool-selection errors, argument or schema mismatches, missing-tool-call cases, multiple errors, and genuine extra tool calls. In particular, when a required tool call is missing, the step receives label 1 even though no tool-call-token span is available. A true ground-truth-exhaustion sentinel is a dataset marker rather than an operational action and is excluded from both training and evaluation. However, an unnecessary tool call emitted after the reference tool-call sequence is exhausted is retained as an operational error.

\subsection{Compared signals}
We compare four scores oriented so that larger values indicate greater estimated error risk: hidden-state probe, mean tool-call-token entropy, negative mean tool-call-token log-probability, self-consistency uncertainty. 
The four-signal head-to-head comparison is restricted to tool-call-bearing records with valid tool-call-token boundaries and finite values for all four signals. Missing-tool-call cases are retained in the full OOF population but excluded from this comparison because they have no tool-call tokens.

\subsection{Metrics}
We report AUROC (which relies on the pairwise ranking of positive and negative examples) and AP (which relies on the frequency of positive labels). We calculate uncertainty by repeatedly resampling tasks and recalculating the metric. We also keep all observations from each sampled task together. The resulting bootstrap intervals reflect task-to-task variation. A negative value favors the baseline since paired detector differences are defined as $\Delta=M_{\mathrm{probe}}-M_{\mathrm{baseline}}$.

\section{Experimental Setup}
\subsection{Benchmark and models}
We use four suites from AgentDojo v1.1.1: Workspace, Banking, Travel, and Slack. We evaluate Qwen2.5-7B-Instruct, Qwen2.5-14B-Instruct, and GLM-4.6V-Flash (9B) in the experiments. We test four scenarios: benign, prompt-injection, system-override, and tool-poisoning. We use non-defended trajectories in all detection experiments. We deploy the detector offline using saved hidden states and uncertainty values. The outputs used for self-consistency are generated during trajectory collection and stored before evaluation, without generating any additional samples from the LLMs during five-fold evaluation.

The trajectory setting consists of 8-bit quantization, 0.0 temperature, a maximum of eight tool steps, and a maximum of 512 generated tokens. For the self-consistency signal, we use four generated outputs for each evaluated step. The primary output is generated with a temperature of 0.0, while three additional outputs are generated with a temperature of 0.7 and top-p of 0.9. The generated tool actions are compared using their function names and arguments while preserving the order of tool calls.  The argument dictionaries are serialized as canonical JSON with sorted keys. A common no-tool signature is assigned to generations without a tool call. We define the agreement score as the ratio of outputs matching the most common tool-call signature. Consequently, self-consistency uncertainty is equal to the complement of the agreement score. The generated outputs and agreement scores are stored during trajectory collection and used for offline evaluation without generating additional samples. We test all three models with the same benchmark set.

\subsection{Hidden-state extraction and probe construction}
At each generation step, a single-token representation is obtained based on the output of the last transformer layer of the model. For obtaining the pre-action state, the hidden state of the last prompt token before generation of tool-call tokens is selected. No pooling operation is applied, and the feature dimension is defined by the hidden size of the model.

\subsection{Five-fold grouped OOF evaluation}
Grouping is performed at the level of suite-task pairs, which are identified as follows: $g=\texttt{suite::task\_id}$. In our experiments, we employ five folds with seed equal to 42. All records belonging to the same group, across steps and conditions, are assigned to the same fold. The fold assignment file is identical for all three models. The five held-out fold sizes are 20, 19, 19, 19, and 19 groups. Each held-out group is scored by a probe trained without that group. A total of 1,027 eligible benign training records are included in the five-fold run across the three models. 

The classifier uses a pipeline containing StandardScaler followed by logistic regression. The standardization parameters are fitted only on the training-task partition of each fold and are applied to the corresponding held-out records. The process requires at least three training task groups for a step-specific fit. During model fitting, balanced class weights and task-balanced sample weights are applied, and logistic regression is run with inverse regularization parameter $C = 1.0$, the \texttt{liblinear} solver, and 300 maximum iterations.

For task steps in a cross-validation fold that do not contain enough training tasks or have only one training class, the classifier process uses a pooled fallback model on only the benign records from the training task groups of the current fold. This run generated 3,763 step-specific predictions and 384 pooled fallback predictions. No predictions were skipped.

\subsection{Population accounting}
The results include 4,147 OOF step records across the three models. The non-defended population used in the process has 4,918 step records. 771 true exhaustion sentinels are excluded from the test population. The fold manifest has 96 task-suite groups, while the incomplete \texttt{slack/user\_task\_11} task-suite group is not available for evaluation.

Of the 4,147 OOF step records, 3,836 are tool-call-bearing records with valid tool-call-token boundaries and finite values for all four detector signals. The remaining 311 records are missing-tool-call cases with no tool-call-token boundaries. They have label 1 and are retained in the full OOF population but excluded from the four-signal comparison. Their tool-call-token entropy and negative mean tool-call-token log-probabilities are undefined.

\subsection{Statistical inference}
Using 10,000 task-cluster bootstrap resamplings, we obtain pointwise intervals for the AUROC and AP metrics as 95\% confidence intervals. 10,000 cluster-swap permutations are used for each comparison in the main paired test. In each permutation, the probe and baseline scores are randomly exchanged for all records within each task cluster, and the paired metric differences are recalculated.  There are 72 comparisons (3 models, 4 conditions, 2 metrics, 3 baselines). We apply Holm correction separately to the AUROC and AP families. As a sensitivity analysis, we report global max-$|t|$ simultaneous intervals from 10,000 task-cluster bootstrap replicates. For these intervals, task groups are resampled with replacement. The same resampled task groups are used across all 72 comparisons within each bootstrap replicate, and the maximum absolute standardized statistic is used to construct the simultaneous confidence intervals. The two procedures differ, and we report their significance counts separately.

\begin{table*}[t]
\centering

\ifdefined\isIEEE
\caption{AUROC and AP for each detector on the common complete-case population. The $n$ column gives the number of tool-call-bearing records with valid tool-call-token boundaries and finite values for all four signals. Each cell reports AUROC/AP. Entropy denotes mean tool-call-token entropy.}
\label{tab:main}
\fi

\small
\setlength{\tabcolsep}{3.0pt}
\renewcommand{\arraystretch}{1.13}
\begin{tabular}{llrcccc}
\toprule
Model & Condition & $n$ & Hidden probe & Entropy & $-\mathrm{MeanLogP}$ & Self-consistency \\
\midrule
\multirow{4}{*}{GLM 9B} & Benign & 303 & 0.699/0.807 & 0.718/0.850 & 0.723/0.856 & 0.650/0.785 \\
 & Prompt injection & 307 & 0.676/0.783 & 0.723/0.846 & 0.725/0.848 & 0.650/0.768 \\
 & System override & 270 & 0.621/0.706 & 0.711/0.796 & 0.720/0.807 & 0.742/0.799 \\
 & Tool poisoning & 323 & 0.711/0.827 & 0.733/0.864 & 0.738/0.869 & 0.721/0.841 \\
\midrule\multirow{4}{*}{Qwen 7B} & Benign & 315 & 0.621/0.778 & 0.799/0.916 & 0.796/0.917 & 0.704/0.844 \\
 & Prompt injection & 327 & 0.642/0.803 & 0.819/0.932 & 0.819/0.934 & 0.750/0.880 \\
 & System override & 282 & 0.575/0.694 & 0.782/0.871 & 0.777/0.864 & 0.645/0.754 \\
 & Tool poisoning & 331 & 0.642/0.803 & 0.799/0.927 & 0.796/0.926 & 0.726/0.870 \\
\midrule\multirow{4}{*}{Qwen 14B} & Benign & 337 & 0.695/0.741 & 0.736/0.793 & 0.729/0.783 & 0.677/0.741 \\
 & Prompt injection & 380 & 0.702/0.815 & 0.763/0.870 & 0.759/0.863 & 0.718/0.817 \\
 & System override & 309 & 0.431/0.573 & 0.758/0.821 & 0.744/0.798 & 0.730/0.781 \\
 & Tool poisoning & 352 & 0.731/0.831 & 0.759/0.865 & 0.756/0.857 & 0.673/0.791 \\
\bottomrule
\end{tabular}

\ifdefined\isACL
\caption{AUROC and AP for each detector on the common complete-case population. The $n$ column gives the number of tool-call-bearing records with valid tool-call-token boundaries and finite values for all four signals. Each cell reports AUROC/AP. Entropy denotes mean tool-call-token entropy.}
\label{tab:main}
\fi

\end{table*}

\section{Main Results}

\subsection{Detector performance across operating conditions}
Table ~\ref{tab:main} presents the AUROC and AP results for all four detectors on the common complete-case population of tool-call-bearing records with valid tool-call-token boundaries and finite values for all four signals. The sample size varies across model-condition pairs. 

Negative mean tool-call-token log-probability, tool-call-token entropy, and self-consistency uncertainty obtain higher AUROC and AP values than the hidden-state probe across all four settings for the Qwen2.5-7B-Instruct model. The largest performance change occurs under the system-override condition. Tool-call-token entropy yields an AUROC of 0.782 and AP of 0.871, while the hidden-state probe yields an AUROC of 0.575 and AP of 0.694.

The results for GLM-4.6V-Flash show smaller performance differences between the detectors. Tool-call-token entropy and negative mean tool-call-token log-probability achieve higher AUROC and AP values than the hidden-state probe across all four conditions. In the system-override condition, self-consistency uncertainty achieves the highest AUROC (0.742). But negative mean tool-call-token log-probability achieves the highest AP (0.807). In general, the hidden-state probe fails to consistently outperform the scalar uncertainty metrics. 

For Qwen2.5-14B-Instruct, the hidden-state probe obtains higher AUROC than the Qwen2.5-7B-Instruct probe under benign, prompt-injection, and tool-poisoning conditions. This pattern does not continue under system-override. In this condition, the hidden-state probe obtains 0.431 AUROC and 0.573 AP, while tool-call-token entropy reaches 0.758 AUROC and 0.821 AP. Negative mean tool-call-token log-probability and self-consistency uncertainty also obtain higher scores than the probe. Therefore, the results do not show a uniform improvement in probe performance as model size increases.

\subsection{Paired detector comparisons}
The paired statistical tests identify six significant differences after Holm correction. All six involve Qwen2.5-14B-Instruct under system-override and favor the scalar uncertainty metrics. Under system-override, the difference between the hidden-state probe and tool-call-token entropy is $-0.3269$ for AUROC and $-0.2486$ for AP ($p_{\mathrm{Holm}}=0.0036$ for both). Comparison with negative mean tool-call-token log-probability yields differences of $-0.3128$ for AUROC and $-0.2252$ for AP (both $p_{\mathrm{Holm}}=0.0036$). The probe--self-consistency differences are $-0.2986$ for AUROC and $-0.2081$ for AP (both $p_{\mathrm{Holm}}=0.0036$). None of the other comparisons remain significant after Holm correction.

\begin{figure*}[t]
  \centering
  \includegraphics[width=0.98\textwidth]{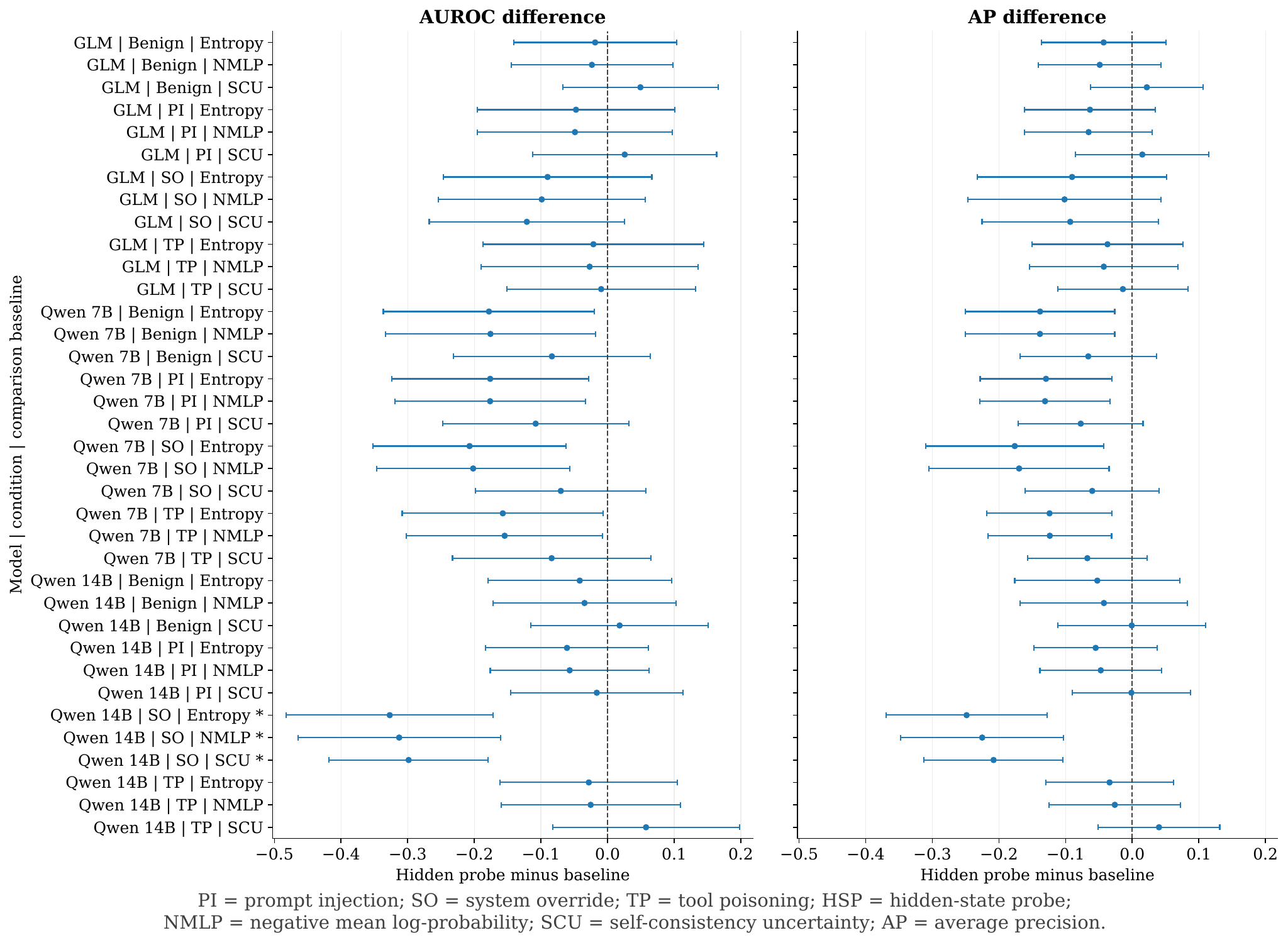}
  \caption{Paired detector differences. Positive values favor the hidden-state probe; negative values favor the scalar comparator. The figure shows family-wise max-$|t|$ simultaneous bootstrap intervals and Holm-adjusted paired cluster-swap permutation results. Asterisks indicate comparisons that remain significant after Holm correction ($p_{\mathrm{Holm}} < 0.05$). Entropy denotes mean tool-call-token entropy.}
  \label{fig:paired}
\end{figure*}

The additional max-$|t|$ bootstrap analysis finds 22 of the 72 simultaneous intervals excluding zero. This result differs from the six significant permutation tests because the two procedures use different approaches to estimate uncertainty. We report them separately rather than treating their results as interchangeable. Figure~\ref{fig:paired} presents the paired differences and simultaneous bootstrap intervals.

\section{Sensitivity Analyses}
We ran two sets of pipeline-level sensitivity analyses using the same seed-42 folds and trajectory caches. They include analyses of data qualification and prediction mode but exclude analyses of transformer ablations, hidden-state pooling, and probe removal.

\subsection{Fallback regime: pooled vs. step-specific probes}
Among the total of 4,147 OOF predictions, 3,763 (90.7\%) are step-specific, and 384 (9.3\%) are pooled across steps. The fallback predictions are not distributed evenly across models. GLM has 161 pooled-fallback predictions, Qwen2.5-7B-Instruct has 195, and Qwen2.5-14B-Instruct has 28. We include all 4,147 predictions in the main hidden-probe evaluation, and we omit no records.

As a sensitivity test, we contrast the distribution for the population with fallbacks with that for the population with step-wise estimates only. For Qwen2.5-14B-Instruct under system override, we have hidden-probe AUROC/AP values of 0.431/0.573 and 0.428/0.564 with and without fallbacks, respectively. Tool-call-token entropy scores change from 0.758/0.821 to 0.762/0.819, negative mean tool-call-token log-probability scores change from 0.744/0.798 to 0.747/0.794, and self-consistency uncertainty scores change from 0.730/0.781 to 0.731/0.778. The qualitative difference between the hidden-state probe and the scalar uncertainty metrics holds even after we discard the fallback estimates.

\subsection{Missing Tool Call Records and Tool-Call-Token Availability}
We classify the 311 records without tool-call-token entropy and negative mean tool-call-token log-probability values as \texttt{missing\_tool\_call} cases. In these records, the model emitted no tool call (\texttt{action\_bearing=False}), so we could not identify a valid beginning and end of a tool-call token sequence (\texttt{action\_boundary\_valid=False}). Since these cases have no tool-call tokens, we keep them in the OOF dataset but do not use them in the four-signal comparison. We do not impute these missing values.

\section{Discussion}
The findings offer a tentative response to the research question. In the AgentDojo dataset, the hidden-state score is not reliably better than uncertainty metrics. This applies most strongly to Qwen2.5-7B-Instruct, where tool-call-token entropy, negative mean tool-call-token log-probability, and self-consistency uncertainty have higher AUROC and AP values than the probe across all four conditions. However, the GLM case is more variable, with no pairwise difference remaining significant after Holm correction and smaller differences between the detectors.

The new Qwen2.5-14B-Instruct results show that model size alone is not enough to determine whether a benign-trained probe can generalize to adversarial settings.  The clearest disadvantage for the probe occurs under system-override, where all three scalar baselines perform better.

The fallback experiment further shows that some OOF predictions arise from a pooled model rather than a step-specific probe. In total, 9.3\% of predictions use the pooled fallback regime. The step-specific-only sensitivity analysis preserves the large difference between the Qwen2.5-14B-Instruct probe and the scalar baselines under system override. Hence, the main result is not explained only by the use of pooled fallback estimates.

The present study further shows that exposure detection and action-correctness detection must not be confused. A hidden representation can carry knowledge of the adversarial environment but may still rank incorrect actions higher than correct actions. This point becomes especially important in light of the recent research on latent prompt-injection-based exposure signals and recognition-action gaps  \citep{dong2026latent}. The current experiment detects tool-call correctness directly, but the reason for the change in discriminative behavior remains unknown.

Finally, a detector ranking is not synonymous with a positive blocking strategy. The present research focuses on analyzing trajectories stored without any defense and cannot validate whether blocking improves task usefulness or reduces the number of attacks based on five-fold detector rankings. This is a question for a separate study.
\section{Limitations}

\paragraph{Scope and Data Splitting} We evaluate three models on a single version of AgentDojo benchmark that covers 96 of 97 expected task groups. Task coverage is limited by the missing Slack task group, which is not available for evaluation. The group split is shuffled $K$-fold rather than stratified group $K$-fold, and the number of independent task groups remains small, which may introduce some variance in our cross-validation folds.

\paragraph{Methodological Constraints} A notable data constraint is that 311 missing-tool-call cases lack valid tool-call-token boundaries. Therefore, tool-call-token entropy and negative mean tool-call-token log-probability values cannot be calculated for these records. Thus, pairwise detector comparisons depend on records with valid tool-call token boundaries, which may affect the comparisons. In addition to that, 384 of our OOF predictions relied on pooled fallback rather than a step-specific probe.

\paragraph{Deployment and Generalizability} Finally, our evaluation was done offline based on pre-stored non-defended trajectories. This paper does not consider the limitations of the online system such as latency, computational expense, and usefulness of blocking the incorrect tool call before execution. Our labels evaluate whether the operational tool action is correct or not. Thus, the above results should not be generalized to natural language hallucinations or agent safety issues.
\section{Conclusion}
In this study, we explored whether hidden-state probes trained on benign tool conditions can identify incorrect behaviors under adversarial conditions. We ran experiments on 96 task groups from AgentDojo across four operating conditions in Qwen2.5-7B-Instruct, Qwen2.5-14B-Instruct, and GLM-4.6V-Flash. The results do not show that the hidden-state probe consistently outperforms entropy, negative mean tool-call-token log-probability, or self-consistency uncertainty. The most significant difference occurred for the Qwen2.5-14B-Instruct model in the case of the system override condition, when all baseline scalar methods gave higher AUROC and AP results and six comparisons turned out to be significant even after Holm correction. The results from Qwen2.5-14B-Instruct do not demonstrate a consistent improvement in hidden-state probe performance over Qwen2.5-7B-Instruct. These results demonstrate that treating hidden-state probes as reliable monitors will require a much more careful approach to how we evaluate probe transferability, training eligibility, and scoring methods before relying on them.

\balance
% All entries in references.bib are the selected 20-paper bibliography.\n\nocite{*}
\bibliographystyle{IEEEtran}
\bibliography{references}

\end{document}